# Intramolecular distances and form factor of cyclic chains with excluded volume interactions

Ana M. Rubio [a], Gabriel Álvarez [b], Juan J. Freire [c],*

[a] *Departamento de Química Física, Facultad de Ciencias Químicas, Universidad Complutense, 28040 Madrid, Saín*

[b] *Departamento de Física Teórica II, Facultad de Ciencias Físicas, Universidad Complutense, 28040 Madrid, Spain*

[c] *Departmento de Ciencias y Técnicas Fisicoquímicas, Facultad de Ciencias, Universidad Nacional de Educación a Distancia, 28040 Madrid, Spain*

\* Correspondence author. Tel/Fax: +34 913988627.

E-mail address: jfreire@invi.uned.es (J.J.Freire).

Short Title: Form factor of a cyclic chain.

**Abstract**

Numerical simulations are performed for isolated cyclic chains with excluded volume. Data are reported for the form factor, $S(x)$, where $x$ is the reduced scattering variable, and also for averages and distributions of the distance between intramolecular units. The averages of distances are compared with two alternative expressions describing their dependence with the number of segments separating the units. The distribution function results are compared with the des Cloizeaux form. Finally the $S(x)$ data are compared with theoretical functions also derived from the des Cloizeaux expression for the distribution function. Moreover, the low $x$ and asymptotic expansions of these functions are obtained. Based on these expansions, simple formulas are proposed to give a good description of the simulation data in the whole range of values of $x$. A comparison with similar results for linear chains is also included.

## 1. Introduction

Both synthetic and natural cyclic polymers are common and important types of molecules[1]. For instance, it is known that DNA may exist in form of ring molecules. Moreover, the conformational properties of ring chain molecules have a special interest given their translational invariance along the chain contour which eliminates the end effects present in linear chains.

The intrachain scattering factor, or form factor, is related to different scattering experiments, and provides a good description of the conformational behavior of chain molecules. The form factor of a flexible polymer chain with $N$ chain units is defined as

$$S(q) = N^{-2} \sum_{j}^{N} \sum_{k}^{N} \langle \exp[i\mathbf{q} \cdot (\mathbf{R}_j - \mathbf{R}_k)] \rangle, \qquad (1)$$

$\mathbf{q}$ is the wavevector that describes the momentum transfer in the scattering, $\mathbf{R}_j$ and $\mathbf{R}_k$ are the position of the $j$-th and $k$-th chain units and $\langle \rangle$ denotes an equilibrium average over the different orientations and the different conformations of the chain. Although $S(q)$ formally depends on vector $\mathbf{R}_j - \mathbf{R}_k$ in Eq. (1), a general orientational average shows that the relevant conformational information needed to evaluate the form factor is the distribution of distances between pairs of units[2]. For a long and flexible polymer, the form factor can be expressed in terms of variable $x=q^2<S^2>$, where $<S^2>$ is the mean quadratic radius of gyration of the chain. At very low $x$, $S(x)$ is similar for all types of chains. However, the form factor behavior is significantly different for different chain models at moderate or large values of $x$.

The form factor of a long ideal cyclic chain with a Gaussian distribution of intramolecular distances is described by the following equation, derived time ago by Casassa [1,3]:

$$S(x) = (2/x)^{1/2} e^{-x/2} \int_0^{\sqrt{x/2}} dt\, e^{t^2} \qquad (2)$$

This expression is valid for $qb \ll 1$, where $b$ is the length of a polymer unit. This implies not very large $x$ since $<S^2>=Nb^2/12$ for cyclic chains in this particular case of "Gaussian" or "unperturbed" chains [1]. For greater values of $q$ or $x$, the scattering experiment probes distances for which the structural details of the units are relevant. As long as the restriction $qb \ll 1$ holds, $x$ may have any value from zero to infinity.

However, excluded volume effects have to be introduced in order to describe the general behavior of any isolated long flexible polymer chain immersed in a good solvent [4]. The intrachain distribution function and averages show large deviations from the Gaussian form, affecting to both the average radius of gyration and the form factor. In the case of cyclic chains, deviations from the Casassa function behavior are expected at high $x$, even for $qb \ll 1$. Some theoretical work has been devoted to give a precise description of $S(x)$. A simple scheme, modifying the intrachain distance form for the averages to take into account excluded volume effects but maintaining their Gaussian distribution, has been proposed by Bensafi et al. [5]. Moreover, a field-theoretical method was applied by Calabrese et al. to obtain the form factor and distribution function of intrachain distances for cyclic chains with excluded volume interactions [6]. This description provides complex formulas and shows that the form of the distribution function is similar to that a generic function previously proposed for a linear chain with excluded volume.

Actually, the case of linear chains has received considerable attention. Time ago, Mazur et al. [7] calculated $S(x)$ for linear chains with excluded volume at high $x$ by using an empirical form for both the mean quadratic intramolecular distances between units and the end-to-end distance distribution function. The results were not completely satisfactory since the distribution of intramolecular distances does not have exactly the same functional form as the end-to-end distance distribution function. Renormalization group and scaling theory has been applied to the calculation of distribution functions of distances. The result in three dimensions can be conveniently written in the des Cloizeaux form [8,9] which is formally equivalent to the function employed by Mazur et al.,

$$P(R_{jk}) = K^{\theta+3} \left(R_{jk} / \langle R_{jk}^2 \rangle^{1/2}\right)^{\theta} \exp\left[-\left(K R_{jk} / \langle R_{jk}^2 \rangle^{1/2}\right)^t\right] \frac{t \langle R_{jk}^2 \rangle^{-3/2}}{4\pi \Gamma[(3+\theta)/t]} \quad (3)$$

where $t = 1/(1-v)$ and $K$ is a normalization constant

$$K = \left\{\Gamma[(5+\theta)/t] / \Gamma[(3+\theta)/t]\right\}^{1/2}. \quad (4)$$

and $\Gamma(a)$ is the Gamma function.

Parameter $v$ is actually a critical exponent, whose value is $v = 0.588$ [4]. Assuming that both units $j$ and $k$ are in the interior part of the chain (as it is always the case in cyclic chains) the numerical value $\theta = 0.71 \pm 0.05$ is also obtained [8]. In previous work [10] we have shown that the behavior of $S(x)$ can be obtained from the approach of Mazur et al. but performing the conformational average with the des Cloizeaux form of the distribution function, Eq. (3). The expression derived for $S(x)$ (an integral) is in good agreement with experimental and simulation data. Low $x$ and also asymptotic expansions are simply obtained from this approach and these expansions are in total agreement with those previously derived using different mathematical approaches [11,12]. The asymptotic limit together with the low $x$ expansion provide Padé approximants that, with a small number of coefficients, are able to describe the exact integral with small error up to the value of $x$ where a few terms of the asymptotic expansion also give good accuracy.

In this work we study the form factor of cyclic chains both from the numerical and theoretical point of view. We provide simulation data for long chains and, extending the theoretical work that we have previously employed for linear chains, we derive an expression for $S(x)$. With this end, we have to make assumptions on the precise form of the distribution function and averages of the intrachain distances in a cyclic ring. These conformational properties are also obtained from our simulation. The comparison of the theoretical expressions with the simulation data of these conformational properties is, therefore, particularly useful to evaluate the validity of the different assumptions needed to obtain $S(x)$.

## 2. Numerical simulations

The model and Monte Carlo algorithms used in our simulations have been described and justified in previous work. The chains have $N$ units whose lengths follow a Gaussian distribution with root mean square $b$ ($b$ is adopted as the length unit). Non-neighboring units interact through a 6-12 Lennard-Jones potential, characterized by the distance and energy parameters $\sigma$ and $\varepsilon$ (the energy unit is the Boltzmann factor $k_BT$). We set the values $\sigma=0.8$ at any temperature-solvent condition. The good solvent or excluded-volume conditions are set with the choice $\varepsilon=0.1$, reproducing the correct behavior in these conditions even for relatively short chains [13].

The algorithm for cyclic chains [14] starts with the generation of a cyclic non-overlapping conformation in a diamond lattice. New conformations are generated from this starting state by choosing two chain units $i$ and $j$ and calculating the two bond vectors $\mathbf{v_i}$ and $\mathbf{v_{j+1}}$ that connect these units to the longest contour in the cyclic chain. Keeping a constant sum $\mathbf{v_i}+\mathbf{v_{j+1}}$ we resample each one of the components of $\mathbf{v_i}-\mathbf{v_{j+1}}$ from a Gaussian distribution with mean equal to zero and mean square deviation $2b^2/3$. This allows us to obtain new positions for units $i$ and $j$. The shorter path of the chain is rotated by an amount defined by random angle $\Phi$ around an axis defined by vector $\mathbf{R_{ij}}$ and then translated to connect again with these new positions. (A similar rotation of a segment of the chain extended from a chosen unit up to its nearest end is applied in the case of linear chains). We compute the total conformational energy in order to accept or reject new conformations, according to the Metropolis criterion.

Typically, we perform 6 runs, each one starting with a different seed number. A run includes the generation of 250,000 conformations for equilibration and 500,000 conformations to evaluate properties, quadratic averages and distribution of intrachain distances and also the form factor obtained from the orientational average of Eq. (1). These properties are stored for every conformation. We obtain arithmetic means over the sample of saved conformations and, finally, we evaluate the final conformational averages as arithmetic means and error bars from the 6 independent runs.

## 3. Theoretical expressions

We assume that the distribution function of intramolecular distances in cyclic chains can be expressed by the des Cloizeaux form, though particular expressions for the average distances have to be considered. This assumption is motivated by the conclusions obtained through the analysis of the field-theoretical derivation of the distribution function by Calabrese et al. [6], whose numerical values were practically coincident with a generic expression equivalent to the des Cloizeaux function. Therefore, we follow the theoretical scheme of our previous work for linear chains, using the analytical procedure outlined in Ref. [7] by Mazur et al., but employing Eq. (3) instead of their empirical expression for the end-to-end distance distribution. However, this scheme has to take into account now the different topology of cyclic chains. First, the double-sum in Eq. (1) is transformed in a single sum over the number of units separating every pair of units in the ring, $n$, from $n=0$ to $N/2$. It is verified that for cyclic chains there are $2N$ equivalent terms of this type covering the whole range of possible pairs of units, except for the special term $n=0$ and the case $n=N/2$ for $N$ even. These exceptions give $\theta(1/N)$ contributions and can be neglected. The single sum over $n$ is then transformed in an integral over variable $n$.

We can express the quadratic average intrachain distances as

$$<R_{jk}^2>/b^2 = f(n), \qquad |j-k|=n \qquad (5)$$

Thus, in the case of unperturbed chains, the averages can be simply written as [1]

$$<R_{jk}^2>/b^2 = pN(1-p) \qquad (6)$$

with $p=n/N$.

Several expressions have been considered to introduce excluded volume effects into Eq. (6). They are inspired by the general formula

$$<R_{jk}^2>/b^2 = (sN)^{2\nu} \qquad (7)$$

which gives the correct result for linear chains with $s=p$ [4]. In the case of cyclic chains, Bensafi et al. [5] proposed a formula that we write as

$$s = [(1/2)-r][(1/2)+r] \qquad (8)$$

with

$$r = n/N - 1/2 \qquad (9)$$

These equations modify the description suggested by Yu et al. [15] that we can write as

$$s^{1/2\nu}=[(1/2)+r^{2\nu}][(1/2)-r^{2\nu}] \tag{10}$$

with

$$r^{2\nu} = (n/N)^{2\nu} - 1/2 \tag{11}$$

Our particular description of $s$ in terms of variable $r$ in Eqs. (9) and (11) allows for more symmetric and easier to handle forms of f($n$). It should be mentioned that the Yu et. al. expression does not hold the formal circularity condition $\langle R_n^2 \rangle = \langle R_{N-n}^2 \rangle$ However, we believe that its validity in the interval $0<n<N/2$ should be solely judged based in its accuracy to describe the chain conformational properties.

Function $S(x)$ can be obtained by integration over variable $u=qR_{jk}$ as

$$S(x) = \frac{t}{\Gamma((3+\theta)/t)y^{3+\theta}} \int_0^\infty du\, sin(u)u^{\theta+1} I_2(u) \tag{13}$$

where $I_2(u)$ has to be previously obtained as an integral over variable $s$. Therefore, we should consider the different definitions for $s$, according to the different options to describe the intramolecular distances. If Eqs. (8)-(9) are employed, $I_2(u)$ is defined as,

$$I_2(u) = \int_0^{1/4} ds\,(1/4-s)^{-1/2}\, s^{-(\theta+3)\nu}\, exp\left[-(u/y)^t\, s^{-\nu t}\right] \tag{14}$$

where

$$y^2 = \frac{\Gamma[(3+\theta)/t]}{\Gamma[(5+\theta)/t]} x \Big/ \int_0^{1/2} p^{2\nu}(1-p)^{2\nu} dp \tag{15}$$

For small values of $x$, we can consider the general small wavevector expansion

$$S(x) = 1 - \frac{x}{3} + \sum_{m=2}^{\infty} (-1)^m b(m) x^m \qquad (16)$$

When $S(x)$ is described by Eqs. (13)-(15) the following convergent expansion around the origin is obtained

$$S(x) = \sum_{m=0}^{\infty} \frac{(-1)^m \, \Gamma[(2m+3+\theta)/t] \, \Gamma(1+2mv)^2}{(2m+1)! \, \Gamma[(\theta+3)/t] \, \Gamma(1+4mv)} y^{2m} \qquad (17)$$

This expression with the numerical values $v=0.588$ and $\theta=0.71$ provides the numerical coefficients $b(m)$ shown in Table 1.

An asymptotic value for $x$ is also obtained,

$$S(x) \sim -\frac{2}{v y^{1/v}} \frac{\Gamma[(3+\theta-1/v)/t]}{\Gamma[(3+\theta)/t]} \Gamma(1/v - 1) \cos\left(\frac{\pi}{2v}\right), \qquad x \to \infty \qquad (18)$$

The numerical value of this asymptotic limit with the exponent values that we have previously used for linear chains, $v=0.588$ and $\theta=0.71$, is 0.615.

On the other hand, when Eqs. (10)-(11) are used for the intrachain distances, the integral over $s$ becomes

$$I_2(u) = \frac{1}{2v} \int_0^{s_0} ds \left[ \frac{1}{2} - \left(\frac{1}{4} - s\right)^{1/2} \right]^{(1/2v)-1} (1/4 - s)^{-1/2} s^{-(\theta+3)/2} \exp\left[-(u/y)^t s^{-t/2}\right] \qquad (19)$$

with

$$s_0 = (1/4)^v \left[1 - \left(\frac{1}{4}\right)^v\right] \qquad (20)$$

and

$$y^2 = \frac{\Gamma[(3+\theta)/t]}{\Gamma[(5+\theta)/t]} x / \int_0^{1/2} p^{2\nu}(1-p^{2\nu})dp \tag{21}$$

$S(x)$ derived from Eqs. (13) and (19)-(21) has the following expansion for low $x$,

$$S(x) = \sum_{m=0}^{\infty} \frac{(-1)^m \Gamma[(2m+3+\theta)/t]}{(2m+1)!\nu\Gamma[(3+\theta)/t]} y^{2m} \beta_{4^{-\nu}}(m+1/2\nu, m+1)$$

(22)

In Eq. (22) the terms contain an incomplete beta function [16] that for this particular case can be expressed as a finite sum,

$$\beta_{4^{-\nu}}(m+1/2\nu, m+1) = \frac{\nu(1-4^{-\nu})^m}{(2m\nu+1)4^{m\nu}} \sum_{k=0}^{m} \frac{(-m)_k}{(m+1+1/2\nu)_k (1-4^\nu)^k}$$

(23)

with

$(a)_0=1$, $(a)_k=a(a+1)\ldots(a+k-1)$

(24)

which, with the numerical values $\nu=0.588$ and $\theta=0.71$, gives numerical coefficients $b(m)$ of Eq. (16) that are also contained in Table 1.

This function also has the asymptotic limit for $x$ given by Eq. (18). However, the numerical values provided by the two approaches are different, because they differ in the numerical relationship between $y$ and $x$, according to Eqs. (15) and (21). For this second approach with $\nu=0.588$ and $\theta=0.71$ we get a limit of 0.700.

**4. Results and discussion**

A first indication of the performance of the different expressions for intrachain distances is provided by computation of the ratio between the mean quadratic radii of gyration of a cyclic and a linear chain $r_c=<S^2>_c/<S^2>_l$ with excluded volume conditions. Renormalization group calculations by Prentis[17] yielded $r_c=0.57$, while some simulations [1,14,18] give a value slightly higher than 0.5 (the unperturbed chain value) for long chains. The present simulations for cyclic and linear chains up to $N=781$ units give the extrapolated value $r_c=0.535\pm0.005$. The simulation data confirm experimental results for polymers in unperturbed (theta) or good solvent conditions, reviewed in Ref. [14] and Ref. [18], that seem to indicate a weak dependence of $r_c$ on solvent conditions. The simple well-known expression related the radius of gyration with the intramolecular distances

$$\left\langle S^2 \right\rangle = N^{-2} \sum_{j}^{N} \sum_{>\ k}^{N} \left\langle \mathbf{R}_{jk}^2 \right\rangle \tag{25}$$

can be transformed to an integral over $n$ or $p$. Using this approach and applying Eq. (7) with $s=p$ for linear chains and Eq. (7) with Eqs. (8)-(9) or Eqs. (10)-(11) for cyclic chains we have obtained numerical values for $r_c$. Eqs. (7)-(9), give a remarkably low value $r_c=0.43$ while Eqs. (7), (10)-(11) lead to the more consistent result $r_c \cong 0.50$.

A direct comparison between theoretical and simulation results for the averages can also be accomplished from our data. In Fig. 1, we compare the theoretical results (with *b* fitted to describe precisely the data corresponding to the shortest |*i-j*|) and our simulation results for a cyclic chain with 781 units. It is clearly observed that Eqs. (7), (10)-(11) give a good description of the simulation data, while Eqs. (7)-(9) show a significant downwards deviation for the highest |*i-j*| and Eq. (6) exhibits a strong and early disagreement, which remarks the poor performance of the Gaussian approximation and the large influence of the excluded volume effects.

Fig. 2 contains the simulation data for two relatively long chains with excluded volume conditions, *N*=246 and 781. The results are presented as generalized Kratky plots [19], $x^{1/2\nu}S(x)$ vs. *x*, which should give a plateau for high *x* in accordance with the predicted asymptotic behavior. This plateau is confirmed by the simulation data (the values of *q* are small enough to prevent the appearance of local model features). Comparing the two values of *N* included in the graphic, it is observed that they show slight differences at intermediate values of *N* for which the asymptotic limit has not been reached. However, the influence of *N* is very small, indicating that the considered chain lengths are near to the long chain limit behavior.

Fig. 3 shows previous simulation data for (shorter) linear chains [20], allowing for a direct comparison of the behavior exhibited by two types of chain topologies. It is observed in Fig. 2 that, in the generalized Kratky representation for cyclic chains, the simulation points show a clear maximum at x≅3. A much flatter (almost undistinguishable maximum) located at higher *x*, was obtained in the case of linear chain. Also, the numerical asymptotic limit for cyclic chains is significantly smaller than for linear chains. These two distinctive features may be of interest for experimental characterization of polymer topology.

Fig. 2 also includes some curves corresponding to different theoretical predictions. It is observed that the Casassa expression for unperturbed chains, Eq. (2) is valid up to $x=2$. However, as expected, this expression cannot describe the high $x$ behavior. Considering the expressions for excluded volume proposed in the preceding Section, defined by Eq. (13) but with different forms for integral $I_2(u)$, both of them give a reasonable description of the simulation data up to $x=2.5$. Therefore they are able to describing points near to the maximum. They also follow the correct qualitative asymptotic behavior. However, when Bensafi et al. intrachain averages are considered, i.e. Eqs. (7)-(9) are used to calculate $I_2(u)$ through Eqs. (14)-(15), we obtain intermediate and asymptotic values significantly smaller than the simulation data. The alternative use of Eqs. (19)-(21) derived from the Yu et al. formula for the intrachain averages, Eqs. (7) and (10)-(11), gives results considerably closer to the simulation data in the whole interval of $x$ values, though some small quantitative differences can still be observed in the intermediate and asymptotic regime (the theoretical curve lies slightly above the simulation points). A similar discrepancy between simulation and theoretical results was observed in the case of linear chains [10,20].

In our discussion of the results for linear chains [10], we conjecture that this difference may be eliminated if a more adequate value of $\theta$ (considered as an empirical parameter) is employed. In fact, some simulation data for the intrachain distance distribution in linear polymers [21] of $N=160$ are apparently more consistent with the value $\theta=0.9$. Following these arguments, we have decided to explore the possibility of using $\theta$=9 in our calculations. In Fig. 4, we show the simulation data obtained for the intramolecular distance distribution function corresponding to cyclic and linear chains with $N$=781. Two cases, $|i-j|$=49 (relatively short value for which, however, we can study a relatively large interval in the short distance range without observing a direct dependence on the intramolecular potential) and $|i-j|$=390$\cong N/2$. It is observed that all the data merge at large distance and that, even at short distances the results for the linear and cyclic chains are very similar and do not exhibit systematic differences. However, there is a remarkable difference between the data obtained for the two values of $|i-j|$. The results corresponding to $|i-j|$=49 are lower and they are better described by a value of $\theta$ close to 0.9. The $|i-j|\cong N/2$ results are, however, above the theoretical line corresponding to $\theta$ =0.71. Therefore, our simulation data seem to indicate that exponent $\theta$ has an empirical dependence with $|i-j|$ which has not been considered in the theoretical approaches.

We have recalculated the theoretical results for $S(x)$ from Eq. (13) and Eqs. (17)-(21) using $\theta$=0.9, since this value appears to be more adequate for short values of $|i-j|$ which gives the more important contribution to the form factor for moderate or high $x$. From Eq. (18) we have obtained a smaller value, 0.676, for the asymptotic limit in excellent agreement with the simulation data. In Fig. 2, we include the results obtained from Eqs. (13) and (19)-(21) with $\theta$=0.9. A good agreement can be observed over the whole $x$ range. The numerical coefficients $b(m)$ of Eq. (16) with this value of $\theta$ are also shown in Table 1. Incidentally, these coefficients are coincident with the values obtained by Calabrese et al. [6], apparently following a totally different theoretical approach that, nevertheless, seems to be practically equivalent.

In Fig. 3, we have included the theoretical values obtained with $\theta$=0.71 and $\theta$=0.9 for linear chains, calculated from Eqs. (5)-(8) in Ref. [10]. A better agreement is clearly found between our previous simulation data for $N$=101, new simulation results obtained in this work for a chain with $N$=781, and the theoretical expression when $\theta$=0.9, which gives the asymptotic limit 1.216. The simulation data lie close to the theoretical line, though slightly below. On the other hand, reliable experimental data of the form factor in a large interval of $x$ values have been reported for linear chains [22] and they show an excellent agreement with the theoretical line in the asymptotic limit (the more pronounced maximum of the experimental data is probably due to particular rigidity effects in the real chains, that are not included in the theoretical model of a totally flexible chain composed of Gaussian units). The low-$x$ expansion coefficients $b(m)$ of Eq. (16) corresponding to this case can be found in Table 1.

The good performance of the results with $\theta$=0.9 for linear and cyclic chains (the latter when used together with the Yu et al. theoretical formulas for intrachain distances) have suggested us to propose simpler numerical formulas in order to compute $S(x)$. Noting the alternating sign pattern of the low $x$ expansions and the fact that the function $S(x)$ decreases almost like $1/x$ for large $x$, it is possible to form Padé approximants with only a few terms to describe the exact expressions up to relatively large values of $x$. In the case of linear chains, an asymptotic expansion was also derived and generic expression were provided [10], see Eqs. (10)-(11) in Ref. [10] ($z$ should read 2 in Eq. (11) of Ref. [10]). Recalculating the results for $\theta$=0.9, the numerical formulas given there are changed to

$$S(x) \cong \frac{1 + 0.0275562x + 0.0115336x^2}{1 + 0.36089x + 0.0528466x^2 + 0.00325559x^3} \qquad \text{for } x<5$$

and

$$S(x) \cong 1.21624/x^{0.8503} + 0.840856/x^{1.7007} - 1.17668/x^{1.9500} - 0.267305/x^{3.1636}$$

$$\text{for } x>5 \qquad (26)$$

with a maximum error of 0.5% at $x \cong 5$ with respect to the numerical integral.

The earlier maximum and technical difficulties in the asymptotic expansion (that it is not so useful in this case because of the occurrence of several terms of similar fractional order) require more coefficients for the Padé approximant in the case of cyclic chains. We find

$$S(x) \cong \frac{1 - 0.140904x + 0.0135922x^2 - 0.000465993x^3 + 8.80468 \times 10^{-6} x^4}{1 + 0.192429x + 0.0169676x^2 + 0.000871902x^3 + 0.0000266987x^4 + 4.01478 \times 10^{-7} x^5}$$

$$\text{for } x<15$$

that, together with the asymptotic limit from Eq. (18)

$$S(x) \cong 0.676/x^{0.8503} \qquad \text{for } x>15 \qquad (27)$$

gives a maximum error of about 1% for $x \cong 15$.

**Acknowledgments**

This work has been partially supported by Grants CTQ2006-06446 and FIS2005-00752 from DGI-MEC, Spain.

Table 1.- Coefficients $b(m)$ of the low $x$ expansions, Eq. (16).

| $m$ | $I_2(u)$, Eq. (17), $\theta=0.71$ | $I_2(u)$, Eqs. (22)-(24), $\theta=0.71$ | $I_2(u)$, Eqs. (22)-(24), $\theta=0.9$ | Linear, Ref.10, $\theta=0.9$ |
|---|---|---|---|---|
| 2 | 0.0599236 | 0.0617016 | 0.0607678 | 0.0777881 |
| 3 | 0.00713984 | 0.00765968 | 0.00737557 | 0.0136197 |
| 4 | 0.000626637 | 0.000705055 | 0.000660936 | 0.0018896 |
| 5 | 0.0000431371 | 0.0000511216 | 0.0000465237 | 0.000215789 |
| 6 | $2.42775 \times 10^{-6}$ | $3.03971 \times 10^{-6}$ | $2.68025 \times 10^{-6}$ | 0.0000208494 |
| 7 | $1.15039 \times 10^{-7}$ | $1.52529 \times 10^{-7}$ | $1.30115 \times 10^{-7}$ | $1.74011 \times 10^{-6}$ |
| 8 | $4.69133 \times 10^{-9}$ | $6.59897 \times 10^{-9}$ | $5.43988 \times 10^{-9}$ | $1.275 \times 10^{-7}$ |
| 9 | $1.67467 \times 10^{-10}$ | $2.5028 \times 10^{-10}$ | $1.99197 \times 10^{-10}$ | $8.3086 \times 10^{-9}$ |
| 10 | $5.30421 \times 10^{-12}$ | $8.43281 \times 10^{-12}$ | $6.47518 \times 10^{-12}$ | $4.8668 \times 10^{-10}$ |

**Figure captions**

Fig. 1. Intrachain distances for a cyclic chain of N=781. Circles: simulation data. Solid line: Eqs. (7)-(9); dash line: Eqs. (7),(10)-(11); dotted line: Eq. (6) with *s=p*, see text.

Fig. 2. Generalized Kratky plot, $S(x) x^{1/2v}$ vs. $x$ ($v = 0.588$) for cyclic chains. Symbols correspond to the Monte Carlo data: x: $N$=246, +: $N$=781. Lines correspond to theoretical results, dotted line, results from Casassa formula, Eq. (2); dotted line: results from Eqs. (13)–(15), $\theta$=0.71; dashed line: results from Eqs. (13), (19)-(21), $\theta$=0.71; solid line: results from Eqs. (13), (19)-(21), $\theta$=0.9.

Fig. 3. Generalized Kratky plot, $S(x) x^{1/2v}$ vs. $x$ ($v = 0.588$) for linear chains. dashed line: theoretical results with $\theta$=0.71 [10]; solid line: theoretical results with $\theta$=0.9. x: Simulation data for a chain of 101 units [20]; +: new simulation data for a chain of $N$=781. Open circles: equation from experimental data proposed by Noda et al. [22]

Fig. 4. Intramolecular distance distributions for a chain of $N$=781. x: |*i-j*|=380, cyclic chains; +: |*i-j*|=380, linear chains; circles: |*i-j*|=49, cyclic chains; squares: |*i-j*|=49, linear chains; solid line: Eqs. (3)-(4) with $\theta$=0.9; dash line: Eqs. (3)-(4) with $\theta$=0.71.

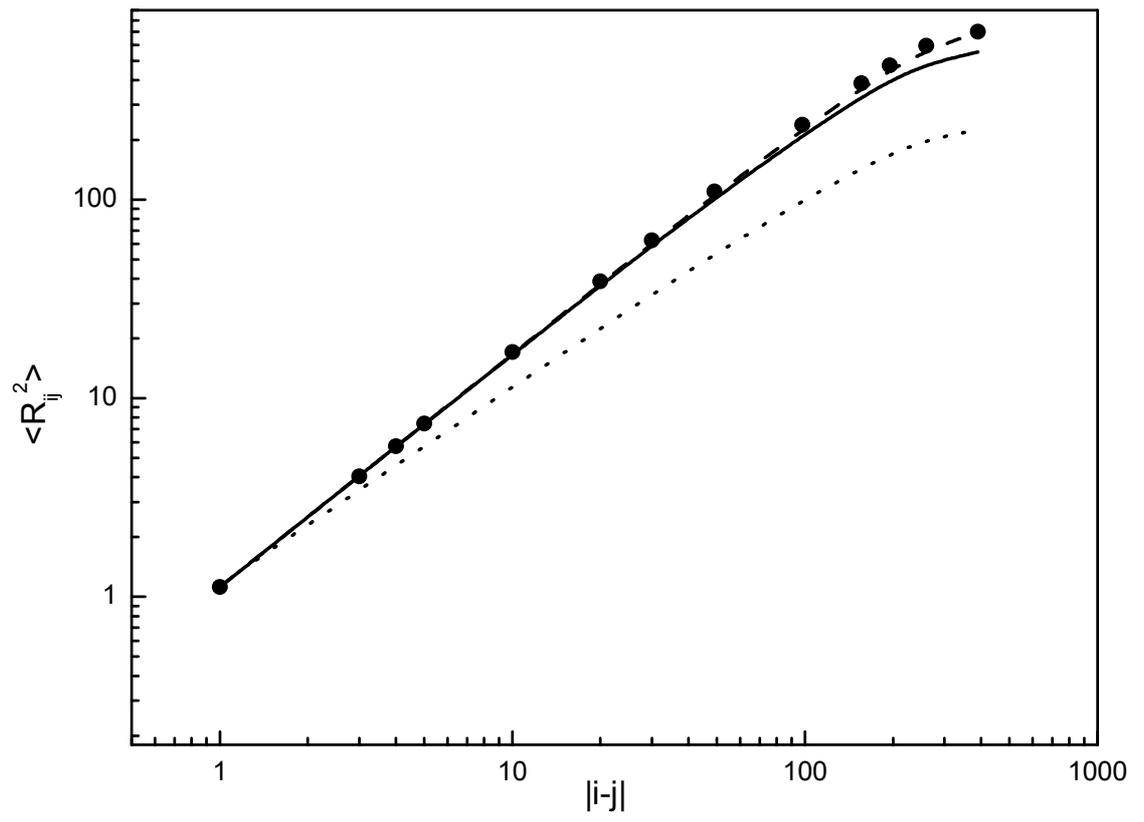

Fig. 1

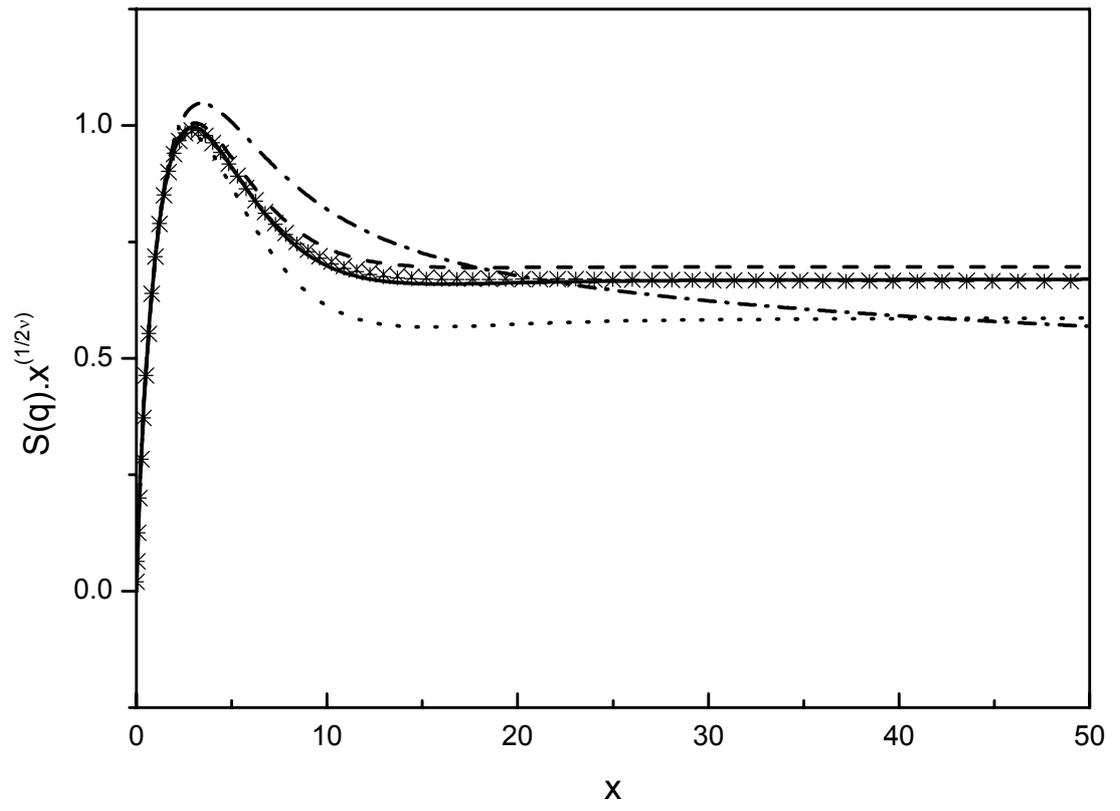

Fig. 2

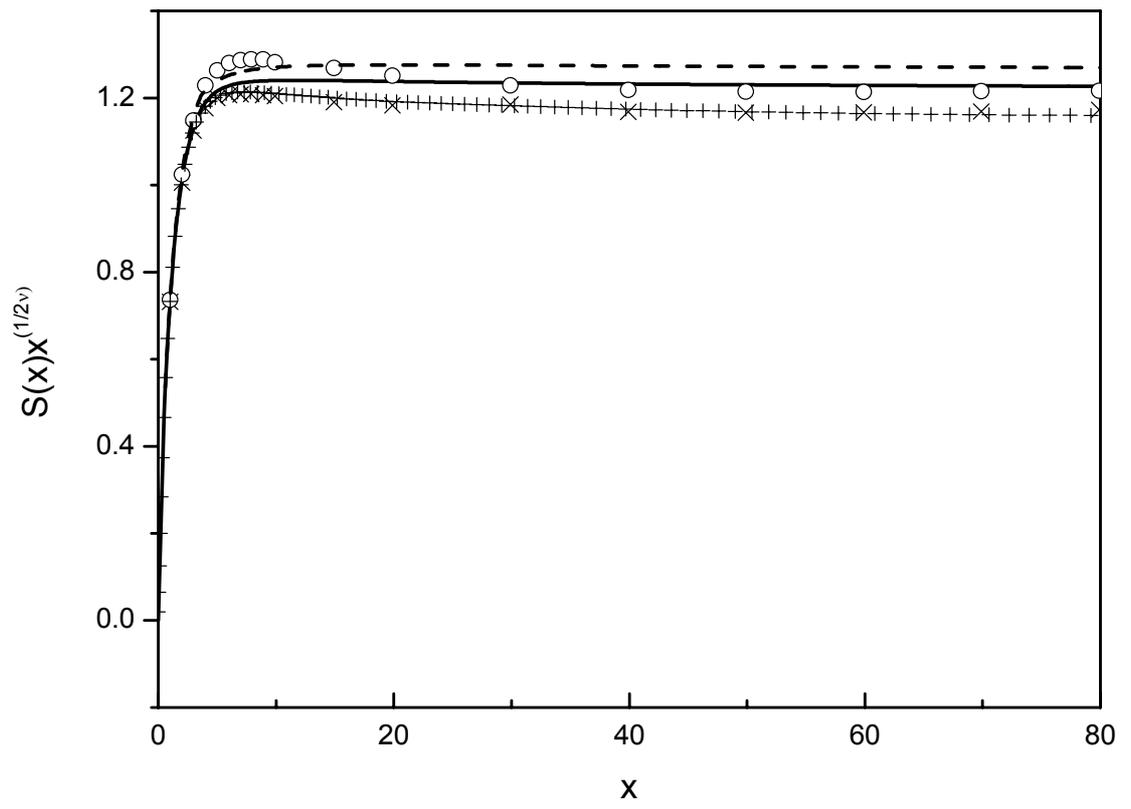

Fig. 3

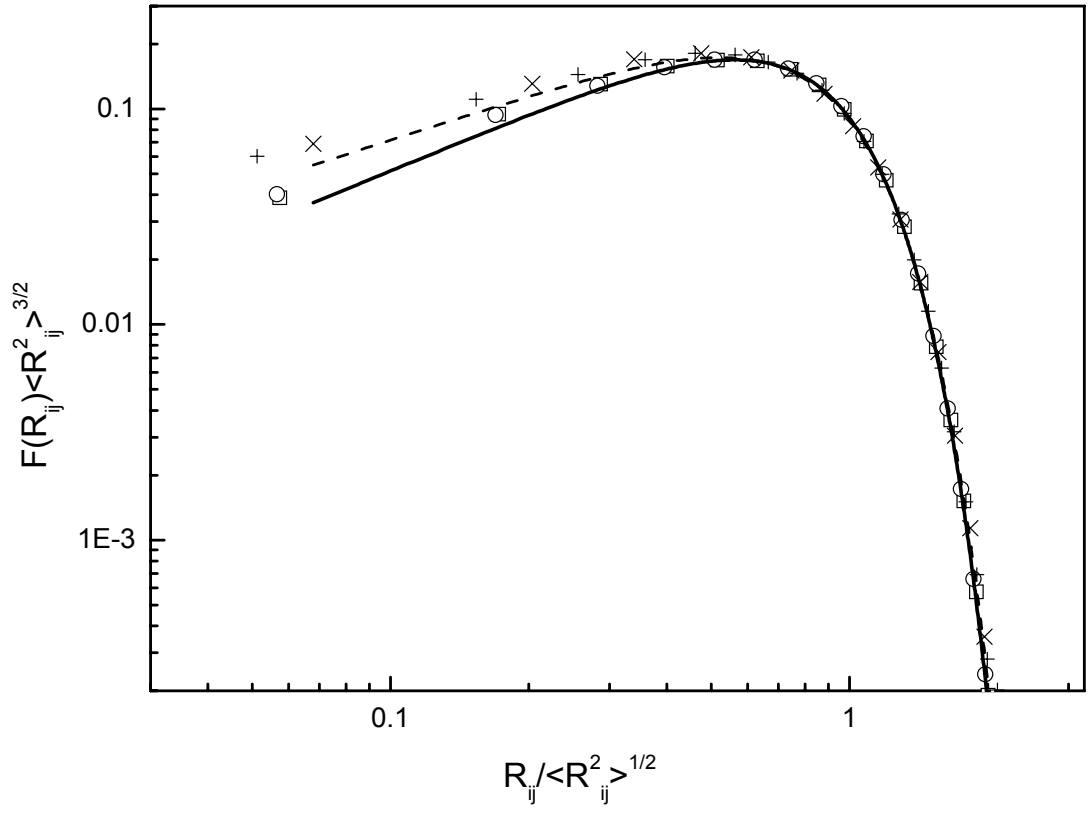

Fig. 4